\documentclass[12pt,preprint]{aastex}

\begin{document}

\def\pdot {\dot P}
\def\Omdot {\dot \Omega}
\def\ltsima{$\; \buildrel < \over \sim \;$}
\def\lsim{\lower.5ex\hbox{\ltsima}}
\def\gtsima{$\; \buildrel > \over \sim \;$}
\def\gsim{\lower.5ex\hbox{\gtsima}}
\def\msole{~M_{\odot}}
\def\mdot {\dot M}
\def\axj {AX~J1845.0$-$0300}
\def\uu {4U~0142$+$61~}
\def\oo {1E~1048.1$-$5937~}
\def\ee {1E~2259$+$586~}
\def\cha {\textit{Chandra~}}
\def\xmm  {\textit{XMM-Newton~}}
\def\xte  {\textit{Rossi-XTE~}}

\title{Pronounced long term flux variability of the Anomalous X--ray Pulsar \oo }
\author{S. Mereghetti, A. Tiengo}
\affil{Istituto di Astrofisica Spaziale e Fisica Cosmica, \\
Sezione di Milano  ''G.Occhialini'' - CNR \\
v.Bassini 15, I-20133 Milano, Italy \\
sandro@mi.iasf.cnr.it}
\author{L.Stella, G.L.Israel, N.Rea}
\affil{INAF, Osservatorio Astronomico di Roma, via
dell'Osservatorio 1, I-00040 Monteporzio Catone, Roma, Italy }
\author{S. Zane}
\affil{Mullard Space Science Laboratory, University College
London, Holmbury St. Mary, Dorking Surrey, RH5 6NT, UK}
\author{T. Oosterbroek}
\affil{INTEGRAL Science Operation Center, Science Operations and
Data System Division, Research and Science Support Dept. of
ESA/ESTEC, P.O. Box 299, NL-2200 AG Noordwijk, Netherlands}

\begin{abstract}
We present    \xmm and \cha  observations of  \oo , being the
first to show evidence for a significant variation in the X-ray
luminosity of  this Anomalous X--ray Pulsar (AXP). While during
the first \xmm (2000 December) and  \cha  (2001 July) observations
the source had a flux consistent with that measured on previous
occasions ($\sim$5$\times$10$^{-12}$ erg cm$^{-2}$ s$^{-1}$), two
more recent  observations found it at a considerably higher flux
level of 2$\times$10$^{-11}$ erg cm$^{-2}$ s$^{-1}$  (2002 August;
\cha) and 10$^{-11}$ erg cm$^{-2}$ s$^{-1}$ (2003 June; \xmm). All
the spectra are fit by the sum of a blackbody with kT$\sim$0.6 keV
and a power law with photon index $\sim$3. No significant changes
were seen in the spectral parameters, while the pulsed fraction in
the 0.6-10 keV energy range decreased from $\sim$90\% in 2000 to
$\sim$53\% in 2003. The spectral invariance does not support the
presence of two physically distinct components in the AXP
emission. The sparse coverage of the data does not permit us to
unambiguously relate the observed variations to the two bursts
seen from this source in the fall of 2001.
\end{abstract}

\section{Introduction}

The Anomalous X-ray Pulsars (AXPs) are a small group of  pulsars
with a rotational period of a few seconds, a fairly stable
spin-down and a very soft X--ray spectrum (Mereghetti \& Stella
1995; van Paradijs, Taam \& van den Heuvel 1995). If they are
neutron stars, as their short spin period suggests, the loss  of
rotational energy inferred from the observed period and period
derivative values is too small to power their luminosity of
10$^{34}$-10$^{36}$ erg s$^{-1}$. AXPs are peculiar because of the
strong evidence, accumulated over several years of intense
observational effort, that they lack a main-sequence or giant
mass-donor companion star (see Mereghetti et al. 2002 for a
review).
This has led to two classes of models involving isolated neutron
stars. In the first class, the emission originates from accretion
of material supplied by  a residual disk (e.g., Ghosh, Angelini \&
White 1997; Chatterjee, Hernquist \& Narayan  2000; Alpar 2001),
while  in the ``\textit{Magnetar}'' model the  energy source is
the decay of an extremely high (10$^{14}$--10$^{15}$ G) magnetic
field (Duncan \& Thompson 1992; Thompson \& Duncan 1995, 1996).
The latter model can explain the properties of the Soft Gamma-ray
Repeaters (SGRs, see Hurley 2000 for a review), a class of
bursting hard X--ray sources with quiescent soft X--ray emission
quite similar to that of the AXPs. The magnetar interpretation has
been supported by the recent discovery of short bursts in two
AXPs, \ee (Kaspi et al. 2003) and \oo (Gavriil, Kaspi \& Woods
2002).

\oo was serendipitously diecovered with the \textit{Einstein
Observatory} as a 6.4 s pulsar near  the Carina Nebula (Seward,
Charles \& Smale 1986) and is one of the best studied AXPs. Early
observations with EXOSAT and GINGA indicated an unusually soft
(compared to accreting binary pulsars) power law spectrum and
measured   a spin-down at a rate of $\sim$1.5$\times$10$^{-11}$ s
s$^{-1}$ (Seward et al. 1986; Corbet \& Day 1990). Higher quality
spectra, not affected by contamination from bright nearby sources,
could be obtained with the imaging instruments on board
\textit{BeppoSAX} (Oosterbroek et al. 1998), \textit{ASCA} (Paul
et al. 2000),  and \xmm (Tiengo et al. 2002). These data showed a
two component spectrum, composed of a blackbody with temperature
kT$\sim$0.64 keV plus a power law with photon index
$\alpha_{ph}\sim$2-3, and a 2-10 keV luminosity of
(1--2)$\times10^{34}$ erg s$^{-1}$  (for an assumed distance d=5
kpc). The source lies at low galactic latitude, resulting in a
significant absorption in soft X-rays (N$_H\sim$10$^{22}$
cm$^{-2}$).

Deviations from a constant spin-down rate were first observed with
ROSAT (Mereghetti 1995). Long term monitoring with \xte carried
out since 1996 (Kaspi et al. 2001) has shown that, compared to
other AXPs for which phase coherent timing  could be obtained over
extended time intervals, \oo is characterized by a relatively high
level of timing noise. The \xte observations were also used to
search for long term flux variability. Unfortunately, the presence
of other X--ray sources in the field of view\footnote{in
particular the highly variable source $\eta$ Carinae
lies at $\sim$40 arcmin} and the uncertainties in the subtraction
of the time variable background, permitted us to use these data
only to measure the pulsed component of the flux. Kaspi et al.
(2001) concluded that the pulsed flux did not vary by more than
25\% in 1996--2000.\footnote{As noted by the same authors, this
does not exclude flux variations, provided that the pulsed
fraction was anticorrelated with the phase-averaged flux in order
to cancel out variations in the pulsed flux.}

On the other hand, \xte detected two bursts of short duration
(about 51 and 2 s respectively)   from the direction of \oo on
2001 October 29 and November 14 (Gavriil et al 2002).
Although the possibility that they originated from a different
source within the instrument field of view cannot be completely
excluded,
it is likely that they were due to \oo, especially in light of the
detection of short bursting activity also in another AXP (\ee,
Kaspi et al. 2003).

Here we present  \xmm and \cha observations showing  significant
long-term variability in the X--ray flux from \oo .

\section{Data analysis and results}

\subsection{\xmm}

\oo was  observed  with \xmm on 2003 June 16  for about 19 hours.
Here we report on the results obtained with the EPIC instrument,
consisting of two MOS  and one PN CCD cameras (Turner et al. 2001,
Struder et al. 2001). After standard data cleaning to remove time
intervals with high particle background, we obtained a net
integration time of 47,000  s in the two MOS cameras, which were
operated in Small Window mode (frame integration time of 0.3 s).
The net exposure was 43,000 s in the PN camera, which was operated
in the standard Full Frame mode with a frame integration time of
73 ms. The medium thickness filter was used in all cameras.

The   data reduction was performed using  version 5.4.1  of the \xmm
Science Analysis System. All the spectra described below were extracted
in the  0.6-10 keV range and rebinned to have at least 30 counts in each energy bin.
Background spectra were extracted from source free regions in the same
CCD chips in which  the source was detected.

\oo had  a net count rate of 3.3 counts  s$^{-1}$ in the PN
camera, more than twice the  value  measured in a short \xmm
observation of 2000 December (Tiengo et al. 2002). For bright
sources, the presence of more than one photon per pixel per frame
time could affect the spectral results. Therefore, we checked for
the possible presence of photon pile-up by comparing the spectrum
extracted from a circular region around the source with that
obtained excluding from the same region the central part (10$''$
radius) where pile-up is expected to occur. Since we found that
the difference in the derived spectral parameters was within their
statistical uncertainties, in the subsequent analysis we used a
circular extraction region of radius 30$''$. The negligible
pile-up in our observation is also confirmed by the fact that the
MOS spectra, which do not suffer from this effect, gave results
fully consistent with the PN spectra (see below).

Single component spectral models (power law, thermal
bremsstrahlung and blackbody) did not provide acceptable fits,
while good results were obtained with the ''canonical'' AXP model
consisting of the sum of a blackbody and a power law. The best fit
parameters (blackbody temperature kT$\sim$0.65 keV, photon index
$\alpha_{ph}\sim$3.3, absorption N$_H$$\sim$1.1$\times10^{22}$
cm$^{-2}$) are similar to those measured on other occasions from
this AXP.

We performed a similar analysis with the two MOS detectors, using
only pattern 0 events. Since the  results were fully consistent
with those obtained with the PN camera, we finally made a  joint
spectral analysis of the three data sets. The corresponding best
fit parameters are given in Table~1, where we also show the
results of a re-analysis, with the updated software and
calibrations, of the 2000 December data.
A comparison of these values clearly shows that,
while the luminosity of \oo  increased by a factor of two, its spectrum
remained virtually unchanged. The ratio of the fluxes in the
two spectral components is, within the errors, consistent with a constant value.

The source pulsations at 6.4 s were easily detected in the 2003
EPIC data. After correcting the PN times to the Solar System
Barycenter, a standard timing analysis based on folding and phase
fitting gave a period  of 6.454835$\pm$0.000001 s. As in previous
observations of \oo, the folded light curve has a single broad
peak of nearly sinusoidal shape. We fitted the background
subtracted folded light curve with a constant (C) plus a sinusoid
of amplitude A. The pulsed fraction, defined as A/C has  values of
46.2$\pm$0.6\% and 56.3$\pm$0.4\%, respectively in the 0.6-1.5 keV
and 1.5-10 keV ranges. These values are significantly smaller than
those observed previously from this source ($\gtrsim$70\%). In
particular, the corresponding values during the 2000 December \xmm
observation are 76$\pm$4\% and 96$\pm$2\%. The time variation of
the source pulsed fraction is illustrated in Fig. 1.

The reduced pulsed fraction, coupled with a larger overall flux,
would suggest the presence of an additional, non-pulsed component
in the 2003  observation. If this were the case, the pulsed flux
and spectrum should be the same in the two observations. We
estimated the spectrum of the pulsed component of each observation
from the difference between the total spectra and the spectra of
the non-pulsed component. The latter were estimated from the phase
interval corresponding to the pulse minimum, for a duration of
10\% of the period.
By fitting the resulting spectra with the blackbody plus power law
model (Table 1) we found that the pulsed flux varied by $\sim$30\%
between the two observations. We thus conclude that a significant
flux variation intervened also in the pulsed signal of the source.

\subsection{\cha}

The \cha satellite observed \oo twice. The first observation was
performed on 2001 July 4 using the High Resolution Camera (HRC-I)
with the main objective of measuring with great accuracy the
source position (Israel et al. 2002). The HRC-I does not provide
spectral information. Assuming the same spectrum as the 2003 \xmm
observation,  the measured count rate of 0.169$\pm$0.005 counts
s$^{-1}$ corresponds to a 2-10 keV observed flux of
$\sim$3$\times$10$^{-12}$ erg cm$^{-2}$ s$^{-1}$. We performed a
timing analysis as described in Israel et al. (2002) determining a
best period P=6.45277$\pm$0.00007 s. This value supersedes that of
Israel et al. (2002), which was affected by an error in the
program used for the Solar System Barycenter correction. The
pulsed fraction, defined as described above,  is 91$\pm$3\%.

The second \cha observation was done using the  ACIS  instrument
in High Energy Transmission Grating (HETG) mode on 2002 August
27-28, for an exposure time of 29,000 s. The data were processed
using CIAO version 2.3. We applied a standard spatial filter and
an order-sorting mask to extract the first--order events. The
exposure map and effective area files
were calculated taking into account  the dithering of the
satellite, and the spectrum (summed over $\pm1$ orders) was
corrected for the effective area, background subtracted and
rebinned so as to have at least eighty photons per bin. These data
have a  time resolution of 3.24 s which does not allow us to
derive a precise period value or to  perform   phase-resolved
spectroscopy. The phase integrated spectrum was well described by
the absorbed blackbody plus power law model with the best fit
parameters reported in Table 1. The corresponding flux of
$\sim2\times10^{-11}$ ergs$^{-2}$ s$^{-1}$, a factor of two larger
than that of the 2003 \xmm observation, is the highest ever
observed from this source.

\section{Discussion}

All flux and period measurements of \oo obtained in the last 11.5
years with imaging X--ray instruments are plotted as a function of
time in Fig. 2. They indicate that, until the 2001 \cha
observation, the flux remained at a relatively stable level
corresponding to a luminosity of $\sim$10$^{34}$ erg s$^{-1}$ (for
d=5 kpc). Comparison of the two \xmm flux values, which are
unaffected by cross-calibration uncertainties related to the use
of different detectors and have small statistical errors, provides
unequivocal evidence for an increased luminosity. Furthermore, the
data from the  2002 \cha pointing suggests that we might have
observed the decaying part of an outburst.

The most recent period value obtained with \xmm  lies   above the
extrapolation of the average spin-down of
$\sim$2$\times$10$^{-11}$ s s$^{-1}$ observed in 1996-2001 (Fig.2
bottom, dotted line). This implies either a period of  increased
spin-down rate ($\pdot >$3.3$\times$10$^{-11}$ s s$^{-1}$) or a
sudden discontinuity ($\Delta$P/P$\sim10^{-4}$) after the 2001
July  \cha observation. Such a high spin-down rate has been seen
on other occasions from this source (e.g. Mereghetti (1995)),
while in the case of a sudden period variation, the sign is
opposite to what is seen in radio pulsar glitches (which typically
involve much smaller fractional variations). We note that a
similar \textit{antiglitch} probably occurred in SGR 1900+14 after
the giant flare of 1998 August 27 (Woods et al. 1999; Palmer
2002).

We cannot rule out that the changes we observed in \oo are related
to the two short bursts detected in the  fall of 2001 (Gavriil et
al. 2002; the time of the bursts is indicated by the vertical line
in Fig.2). In this respect it is interesting to compare our
results with the properties of \ee , the other AXP that showed
short bursting activity (Kaspi et al. 2003, Woods et al. 2003).
\ee emitted more than 80 short bursts in 2002 June. They were much
more similar to those of SGRs than the two bursts seen in \oo,
although their peak luminosity of (1-400)$\times$10$^{36}$ erg
s$^{-1}$ (2-10 keV, assuming isotropic emission and d=3 kpc, Kaspi
et al. 2003) did not reach the largely super-Eddington values seen
in bursts from SGRs. The onset of bursting activity in \ee was
accompanied by an order of magnitude increase of the X--ray flux
and a large glitch ($\Delta\nu/\nu$ = 4$\times$10$^{-6}$). The
subsequent flux decay, after a steep initial phase, was rather
slow, reaching the pre-burst level after about one year. The
spin-down rate for the first weeks after the glitch was about a
factor 2 larger than the pre-glitch value, leading to a partial
recovery of the frequency jump, but the following long term
spin-down rate was only 2\% \textit{smaller} than before the
glitch.

Although the \oo data shown in Fig. 2 (top) are rather sparse, the
flux evolution after the bursts is broadly consistent with that of
\ee. However, the sign of the putative glitch and the long term
spin-down torque indicate significant differences in the frequency
evolution of these two sources. Furthermore, our preliminary
timing analysis of public data from the \xte satellite does not
provide evidence for any major timing discontinuity after the time
of the \oo bursts.
Another difference is the long lasting change in the pulse profile
of \oo. Such a large variation in the pulsed fraction is the first
observed in this source, and if it originated at the time of the
bursts, it persisted for at least 1.5 year. For comparison, in \ee
the variations in the pulse profile were recovered within about
two months from the bursts (Woods et al. 2004).

Luminosity variations are commonly seen in accreting sources and
generally explained as due to changes in the mass accretion rate,
while under the magnetar hypothesis long lasting flux enhancements
are typically related to the onset of crust fractures. For
instance, twists in the external magnetic field induced by large
scale fractures of the crust can force a persistent thermoionic
current through the magnetosphere (Thompson et al. 2000). As a
consequence, ions are lifted off the surface of the neutron star
by their thermal motion, and the counterstreaming electrons are
electrostatically accelerated to bulk relativistic speeds. The
work done on electrons is then released in the form of Comptonized
thermal photons, with a luminosity $L_X \approx 3 \times 10^{35}
\theta A/Z (B/10B_{qed})$~ergs s$^{-1}$, where $\theta$ is the
twist angle and $B_{qed}\sim$4.4$\times10^{13}$~G. The resulting
steady output in particles has been proposed to explain the factor
2-3 increase in the persistent X--ray flux of SGR 1900+14 after
the event of 1998 August 27 (Thompson et al. 2000), and in an
analogous way it may explain an increase of a factor of 4 in the
luminosity of \oo . Since the latter corresponds to
$\sim$3$\times10^{34}$ ergs s$^{-1}$, if B$\sim$B$_{qed}$ the
required twist angle is $\lsim 1$~rad. If the flux increase is due
to a static twist in the surface magnetic field, this will not
necessarily be associated to the substantial increase observed in
the torque, since the current flow is contained well inside the
Alfv\'en radius. However, the enhanced spin-down may be induced by
an increase in crustal fractures and variation in the rate of
fractures induced by persistent seismic activity, the same
mechanism responsible of the high timing noise during
outburst-free epochs.

An alternative possibility, which accounts for both spin-down and
luminosity, is the episodic onset of a wind (Duncan 2000). In this
scenario, frequent small scale fractures in the crust produce
quasi-steady seismic and magnetic vibrations, energize the
magnetosphere and drive a diffuse, relativistic outflow of
particles and Alfv\'en waves. The resulting wind luminosity,
$L_W$, is proportional to the magnetic energy density in the deep
crust, i.e. $\propto$B$_{crust}^2$, with the onset of the wind
being possible only for $B_{crust} < (4 \pi \mu)^{1/2} \sim 6
\times 10^{15}$~G, where $\mu$ is the shear modulus in the deep
crust. For stronger fields, evolving magnetic stresses overwhelm
lattice stresses and the crust deforms plastically instead of
fracturing, turning off the Alfv\'en wind. In principle, this
gives an upper limit $L_W \lesssim 5 \times 10^{36}$~ergs
s$^{-1}$. In reality, as noted by Duncan (2000), magnetar winds
must be mild enough to produce no detectable radio emission, and
$L_W$ is most likely comparable to the steady X--ray luminosity
$L_X$ emitted by the hot stellar surface. When the persistent
luminosity of the source, $L= L_X + L_W$ exceeds the magnetic
dipole luminosity $L_{mdr}$ as calculated from the stellar dipole
field and angular velocity, the spin down torque grows by a factor
of $\sim (L_/L_{mdr})^{1/2}$. In this contest, an episodic
increase of a factor $\sim4$ in luminosity could mean the spin
down rate increased by a factor $\sim2$ (Duncan 2000), as observed
here. The episodic onset of a non-pulsed wind component can also
account for the decrease in the pulse fraction. However, we found
difficult to reconcile this scenario with the spectral invariance
since there is no a priori reason why the spectrum of the wind
component should appear so similar to that of the persistent
X--ray source.


In fact, the \oo spectrum remained unchanged despite the large
variations of the flux and pulsed fraction. As shown above, the
flux difference between the two \xmm observations cannot be
ascribed only to the variation of the non-pulsed component.
Furthermore, the spectral invariance indicates that the unpulsed
flux is not distinguishable, from the spectral point of view, from
the pulsed one. In other words, both the pulsed and non-pulsed
components are fitted by the same blackbody plus power law model.
This supports the idea that the blackbody and power law spectral
components of AXPs do not represent physically distinct emission
processes, as  was already suggested by the small energy
dependence of the AXP pulsed fractions (\"{O}zel  et al. 2001). A
possibility is that the emission is of a thermal nature and the
power-law in the fit simply reflects the inadequacy of describing
a complex neutron star atmosphere with a simple blackbody model.
The luminosity variation could  be attributed to an increase in
the area of the emitting surface, rather than to a temperature
variation; this would also explain the reduction in pulsed
fraction.

\section{Conclusions}

The data reported here provide solid evidence for long term
variations in the  flux and pulsed fraction of the AXP \oo .
Independently of the model adopted to explain the X--ray emission,
the spectral invariance argues against the presence of two
physically distinct spectral components.

The sparse coverage of the data does not permit us to
unambiguously relate the observed variations to the two bursts
seen in the fall of 2001. Furthermore, the differences with
respect to the behavior of \ee after the 2002 June outburst
suggest that the flux variations in \oo are not necessarily
related to SGR-like activity.

Long term flux variations have been reported for the AXP candidate
AX J1845.0--0300 (Vasisht et al. 2000) and for the latest
discovered member of the AXP class,  XTE J1810$-$197 (Ibrahim et
al. 2003, Gotthelf et al. 2004). The latter source was first seen
in a bright state (F$\sim6\times10^{-11}$erg cm$^{-2}$ s$^{-1}$,
2-10 keV) at the beginning of 2003. Its flux then decreased by a
factor $\sim$3 in the first half of 2003. Archival data from ROSAT
and ASCA, in which XTE J1810$-$197 was two orders of magnitude
fainter, indicate that this source behaves like a transient.
Interestingly, XTE J1810$-$197 is quite similar to \oo also
because of its spectral parameters, nearly sinusoidal pulse
profile with  a large pulsed fraction, spin-down at
$\sim$10$^{-11}$ s s$^{-1}$ with considerable timing noise, and IR
counterpart (Israel et al. 2004).

These observations suggest that luminosity variations in AXPs are
more common than previously thought and not necessarily associated
with the emission of energetic flares as in the classical SGRs.

This work has been partially supported by the Italian Space
Agency. Based on observations with \xmm, an ESA science mission
with instruments and contributions directly funded by ESA member
states and USA.


\clearpage

\begin{table*}[h]

\caption{{\it  Results spectral fits}}


 \begin{center}

 \begin{tabular}{lcccccc}

 \hline

\smallskip

  & Absorption          & Photon &  kT & R$_{BB}^{(a)}$ & F$_{PL}^{(b)}$ & F$^{(c)}$     \\

  &(10$^{22}$ cm$^{-2}$)& index & (keV)& (km)         &               &  \\

\hline

\smallskip

2003 Jan & 1.11$\pm$0.02 & 3.32$\pm$0.05 & 0.630$\pm$0.007 & 1.28$\pm$0.03 & 5.0$\pm$0.3 & 10.1$\pm$0.4 \\

2000 Dec$^{(d)}$ & 0.96$\pm$0.09 & 2.9$\pm$0.2 & 0.63$\pm$0.04 & 0.8$\pm$0.1 & 2.8$\pm$0.8 & 4.7$\pm$0.3 \\

& & & & & &   \\

2003 Jun$^{(e)}$ & 1.1$\pm$0.1 & 3.2$\pm$0.2 & 0.67$\pm$0.03 & 0.9$\pm$0.1 & 2.9$\pm$0.9 & 6.2$\pm$0.5  \\

2000 Dec$^{(e)}$ & 1.0$\pm$0.3 & 2.9$\pm$0.7 & 0.61$\pm$0.07 & 0.8$\pm$0.3 & 2.3$\pm$2.0 & 4.1$\pm$0.3  \\

 & & & & & &  \\

2002 Aug$^{(f)}$ & 1.11 (fixed) & 2.7$^{+1.6}_{-0.7}$ & 0.62$\pm$0.07 & 2.1$^{+0.7}_{-0.4}$ & 9$\pm$1 & 22$\pm$2   \\

\hline
\end{tabular}


\end{center}

All errors are at the 90\% confidence level for a single
interesting parameter

$^a$ Radius at infinity for an assumed distance of 5 kpc.

$^b$ Observed flux of the power law component  in units of
10$^{-12}$ erg cm$^{-2}$ s$^{-1}$ (2-10 keV)

$^c$ Total flux (2-10 keV) in units of 10$^{-12}$ erg cm$^{-2}$ s$^{-1}$

$^d$ These values supersede the ones reported in Tiengo et al.
(2002)

$^e$ Pulsed flux only

$^f$ \cha HETG observation

\end{table*}

\clearpage

\figcaption{Folded light curves (0.6-10 keV, EPIC PN ) of \oo
during the December 2000 (left) and June 2003 (right)
observations. The corresponding pulsed fractions (see text for
definition) are 89\%$\pm$1\% and 52.8\%$\pm$0.3\%.
 \label{fig1}}
\epsscale{.8} \plotone{fig1.ps}

\clearpage

\figcaption{Flux and period history of \oo. The vertical dotted
line indicates the time of the bursts seen with \xte . Top panel:
Observed  flux in the 2-10 keV range. Only measurements with
imaging instruments are included.  To derive the first four
values,  we re-analyzed the ROSAT data and converted the
corresponding 0.6-2.4 keV fluxes to the 2-10 keV range by assuming
the same spectral parameters as the 2000 \xmm observation. The
horizontal line is a fit with a constant, excluding the last two
points. Bottom panel: Spin period evolution. The points without a
corresponding one in the top panel are from \xte. The solid line
segments show two phase-coherent solutions from Kaspi et al.
(2001).
 \label{fig2}} \epsscale{.8} \plotone{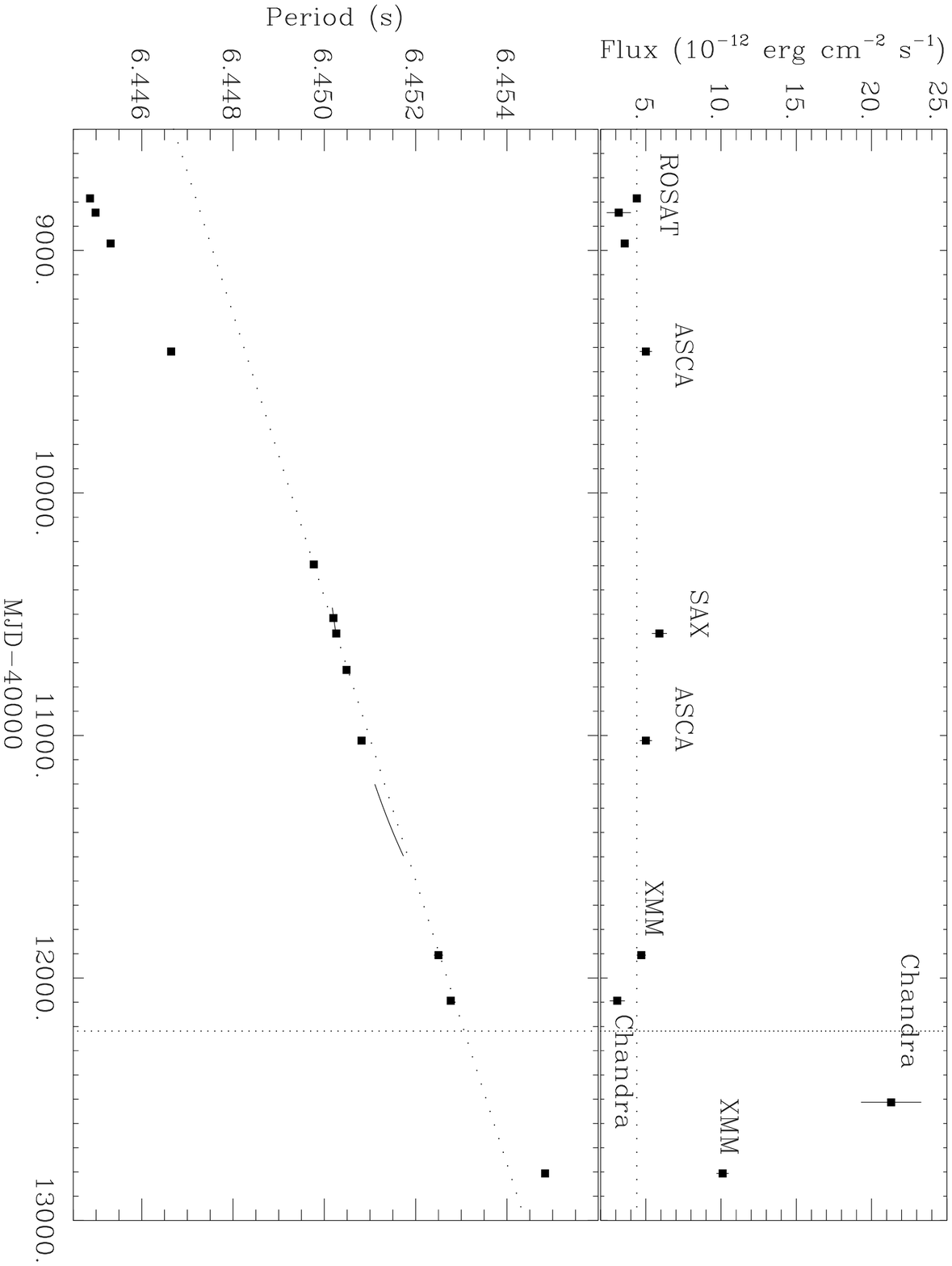}

\end{document}